\documentclass[11pt,notitlepage,nofootinbib]{revtex4-1}

\usepackage{epsf}
\usepackage{latexsym}
\usepackage{amssymb}
\usepackage{amsmath}
\usepackage{bm}
\usepackage[dvips]{graphicx}
\usepackage{array}
\usepackage{xcolor}
\usepackage{soul}

\date{\today}
\def\d3{^{(3)}\nabla}

\begin{document}
\title{Effects of primordial magnetic fields on  21 cm multifrequency angular power spectra}

\author{Kerstin E. Kunze}

\email{kkunze@usal.es}

\affiliation{Departamento de F\'\i sica Fundamental, Universidad de Salamanca,
 Plaza de la Merced s/n, 37008 Salamanca, Spain}

\begin{abstract}
The  redshifted cosmic 21 cm line signal of neutral hydrogen  provides the  possibility to constrain  the matter power spectrum.
Cross correlating  temperature maps at different frequencies   
(corresponding to different redshifts along the line of sight) allows to determine multifrequency angular power spectra.
Primordial magnetic fields raise the amplitude of  the linear matter power spectrum on small scales dominating over the contribution of  the adiabatic 
primordial curvature mode.
Multifrequency angular power spectra could provide a new possibility to probe primordial magnetic fields including its evolution present before decoupling.
Here first multifrequency angular power spectra of the 21 cm line signal are obtained 
for cosmological models including the adiabatic, primordial curvature mode as well as the 
compensated magnetic mode for different values of the magnetic field parameters. 
For this temperature maps are simulated with the modified linear matter power  spectra for magnetic fields with  
magnetic field strength $B_0=4$ nG and spectral indices $n_B=-2.9$ and $n_B=-2.5$, respectively, which have been chosen as examples.
Moreover foregrounds have not been taken into account. 
As examples these  multifrequency angular  power spectra  have been calculated for
frequency ranges  set around central frequencies of uGMRT Band 3 data as well as MeerKAT L band data. For these
using only system  noise 
signal-over-noise ratios are obtained as well as for  SKA1-MID which is part of  the currently under construction SKAO.
Within this simple model results of this first study  seem to be promising for constraining magnetic field parameters especially for SKA1-MID.  
\end{abstract}

\maketitle

\section{Introduction}
\label{s0}
\setcounter{equation}{0}
Effects of a putative primordial magnetic field present since  before decoupling have been studied 
in very different physical settings and epochs of the universe (e.g., \cite{widrow, rev4,Durrer:2013pga,subramanian}).
Cosmological, primordial  magnetic fields are modeled to be stochastic with observations requiring them to be of small amplitudes (cf., e.g., \cite{Planck2015cmf,kk11,sl,pfp,kb,Giovannini:2004aw,KimOlintoRosner}).
These interact with the cosmic plasma leading on the one hand to  sourcing of density perturbations \cite{wasserman,KimOlintoRosner} 
and on the other hand to partial diffusion of their energy in the pre- as well as the post recombination universe \cite{sb,jko,sesu}.
Whereas the former has potentially important effects on large scale structure the latter can influence the thermal and ionization history of the universe as well.
To study these effects in detail requires coupled gravity and MHD simulations such as, e.g., carried out in  \cite{ralegankar}. 
There is a whole range of possible implications  such as early dark halo formation \cite{tashiro2011}, early reionization  \cite{sesu},
increase in star formation rates in dwarf galaxies at early times and formation of mini halos  \cite{sanati2020}
(for an overview and references see e.g. \cite{CruzKamionkowKovetz})

Therefore, generally
speaking primordial magnetic fields can influence the 21 cm line signal  by 
two  types of mechanisms \cite{sesu,kuko15,PaoChlubP18cmf-ioniz}. Dominantly,  the presence of the Lorentz term in the baryon velocity equation  leads to an increase of the amplitude of the total linear matter power spectrum on small scales (cf section \ref{s1}). This is also the case at higher than linear order. However, when non linear effects become important full numerical solutions become necessary and estimates of the resulting power spectra become more difficult (cf. e.g.  \cite{ralegankar}). One important effect, however, can effectively be implemented which is due to damping of primordial magnetic fields by photon viscosity before recombination leading to a cut-off in the magnetic spectrum at the damping wave number (cf. section \ref{s1}).
Subdominantly, magnetic field  dissipation in the post recombination universe by decaying MHD turbulence and ambipolar diffusion can lead to additional heating of matter \cite{CruzKamionkowKovetz}.
Here the focus is on the effect of primordial magnetic fields on the linear matter power spectrum and subsequently on 21 cm intensity maps \cite{kk23,kk21}. 
Moreover, cross correlating these at different frequencies  the angular power spectra are calculated yielding  
multi frequency angular power spectra (MAPS) \cite{datta2007}. This allows to study  the evolution of the signal with varying frequency or redshift due to the light-cone effect, particularly important during the epoch of reionization. However, in general, this leads to the breaking of statistical homogeneity  and thus breaks the ergodicity of the signal.   
This also implies that the spherically averaged 3D power spectrum in Fourier space $P(k)$ no longer contains all the statistical information
\cite{mondal2020,mondal2018}. 
This can be avoided by 
choosing smaller data cubes or equivalently small redshift or frequency intervals (bandwidths) where the evolution along the line of sight (LOS) can be neglected \cite{trottMWA, mondal2020}. 

However, the 21 cm MAPS does not assume statistical homogeneity and thus captures the  full statistical information \cite{mondal2020}.

Cosmological magnetic fields only contribute subdominantly to the total cosmic signal  and thus are   difficult to detect. 
Here the effect of primordial magnetic fields on MAPS will be considered as a possibility  to constrain  magnetic field parameters.

Moreover only the  HI signal will be considered as this is the main focus of  current  and future cosmic 21 cm line observations with instruments such as  the already operational Giant Metrewave Radio Telescope (GMRT) \cite{gmrt}, MeerKAT which is a precursor radio telescope array  of the not yet completed Square Kilometre Array Observatory SKAO \cite{SantosSKA, ska1-cosmo-redbook}. MeerKAT ("more KAT"  (Karoo Array Telescope)) \cite{SantosMeerKAT} will form part of the mid-frequency component of SKA Phase 1 \cite{ ska1-cosmo-redbook} referred to as SKA1-MID. Characteristics of these three instruments will be used here as examples in the numerical solutions.

In general foregrounds and instrument noise make the task of extracting the cosmic signal difficult. However the relevant foregrounds in the frequency range of the cosmic 21 cm signal are rather well studied.  An interesting possibility is to use MAPS as foregrounds are generally spectrally smooth and vary at most very little within frequency intervals \cite{elahi2301.06677,datta2007}. 
There are different foreground removal techniques as well as  numerical programs to obtain, e.g., the cosmological 21 cm line signal as well as angular power spectra. Here the numerical code \texttt{fg\_rm}  \footnote{\texttt{https://github.com/damonge/fg\_rm}} was used. Further details of foreground emission and removal can be found in appendix \ref{app}.

Here the focus is on the effect of a primordial magnetic  field  on the angular power spectra in the theoretical model.
Thus it is assumed that  foregrounds can be fully understood and  their effects removed from the observational data. Including the contribution from a stochastic magnetic field angular power spectra have been obtained from auto correlations as well as in general cross correlations of  21 cm brightness temperature maps at different frequencies. For this purpose  21 cm line signal temperature maps have been simulated using  linear matter power spectra calculated with a modified version of the {\tt CLASS} code \cite{kk21}  (for references of the original {\tt CLASS} code cf  \cite{class1,class2,class3,class4,class5})
for the total adiabatic and magnetic mode for different choices of the magnetic field parameters. Subsequently these have been used as initial matter power spectra  in the  {\tt crime}  numerical code \cite{crime} to obtain the 21 cm brightness temperature maps.

After modifying accordingly \texttt{fg\_rm}  to cross correlate 21 cm signal maps at different frequencies the corresponding  angular power spectra of the auto- and cross correlations have been calculated.
In the following the resulting MAPS are shown.
Prospects of constraining magnetic field parameters are estimated using signal-over-noise ratios. These are calculated taking into account only system noise.
Numerical results are shown for the upgraded Giant Metrewave Radio Telescope (uGMRT) \cite{elahiMNRAS529,pal2208.11063,elahi2301.06677,uGMRT}, MeerKAT   \cite{PaulMeerKAT2023} as well as the prospective SKA1-MID  \cite{ ska1-cosmo-redbook}.   
 
 In the numerical solutions the best fit parameters of the base model derived from Planck 2018 data only are used \cite{Planck-2018}, in particular, $\Omega_{\Lambda}=0.6842$, $\Omega_m=0.3158$, 
$\Omega_bh^2$ = 0.022383, $\Omega_m h^2$ =0.14314, $H_0 = 67.32 $ km s$^{-1}$Mpc$^{-1}$ and for the adiabatic mode
$A_s = 2.101\times 10^{-9}$ and  $n_s = 0.96605$. 

\section{Cosmic signal}
\label{s1}
\setcounter{equation}{0}
$\Lambda$CDM is currently the standard, minimal parameter model to describe the observational universe.
Observations of the temperature anisotropies and polarization of the cosmic microwave background (CMB) 
are best fitted by an initial adiabatic curvature mode (e.g. \cite{pdg2024,Planck-2018}).
Any additional contributions such as due to a primordial magnetic field have to be subdominant.

The observed global isotropy on large scales could be considered as a  motivation for considering a gaussian, random magnetic field.
There are a range of models to generate primordial magnetic fields in the very early universe, broadly divided into two classes, if either magneto-genesis
takes place during inflation or during a phase transition. These models predict stochastic magnetic fields characterized by amplitudes and spectral indices
as well as their correlation scales.
During inflation magnetic fields are generated from quantum fluctuations of the electromagnetic field and subsequent amplification of perturbations on superhorizon scales, similarly to the inflationary mechanism of generating the primordial curvature perturbation. In this case spectral indices are negative and correlation scales do not pose a problem. Most of  the models lead to non helical magnetic fields. 
Magnetic field generation during phase transitions involves causal processes on subhorizon scales and generically turbulent motions resulting in helical magnetic fields  and positive spectral indices. However, in this case correlation lengths are smaller than the Hubble radius at the epoch of generation leading to very small magnetic field strengths on galactic scales. Though correlation lengths could be increased by inverse cascade which takes place for helical magnetic fields
 (for reviews,  cf., e.g., \cite{SubramaRep2016, Durrer:2013pga,rev4}). 
Magnetic fields generated in the very early universe lead to additional contributions to the total density perturbation $\Delta$, anisotropic stress $\Pi$, as well as the evolution equation of the baryon velocity, 
(see e.g. \cite{kk21,sl12,sl,kk11,kk12}).
The latter being the addition of the Lorentz term which couples to the ionized matter part implying, using conformal time and scale factor $a$, 
\begin{eqnarray}
\ddot{\Delta}_m+{\cal H}\dot{\Delta}_m-\frac{3}{2}{\cal H}^2\Delta_m=\frac{\rho_{\gamma}}{\rho_m}k^2L_B,
\end{eqnarray}
where  the total matter perturbation $\Delta_m=R_b\Delta_b+R_c\Delta_c$, $R_i=\frac{\rho_i}{\rho_b+\rho_c}$, i=b(aryon),c(dm) and ${\cal H}=\frac{\dot{a}}{a}$.
The Lorentz term is defined by $L_B=\frac{2}{3}(\frac{1}{3}\Pi_B-\Delta_B)$. In particular, $P_m(k)\sim k^4P_{L_B}(k)$ (for further details see \cite{sl12}).
Thus here it can be seen that the Lorentz  term causes additional power on small scales in the linear matter power spectrum \cite{sl12,sesu, AdiCruzKamion23}. For further discussions of  
effects of primordial magnetic fields on the matter power spectrum including at higher order, see e.g. \cite{ kk22}.

In the standard $\Lambda$CDM model initial conditions are set for numerical calculations of the CMB anisotropies long before photon decoupling.  
As is the case for other contributions to the  total energy density perturbations and anisotropic stress there are different types of initial conditions, e.g. adiabatic or isocurvature, which applies to the magnetic mode as well. Here the compensated magnetic mode which  yields an isocurvature type perturbation is used in the numerical solutions (for further details cf, e.g.,  \cite{sl, pfp,sl,kk11,kk12, Giovannini:2004aw}).

Here  the simplest model is considered so that
the cosmological magnetic field ${\bf B}$ is assumed to be a non helical, gaussian random field determined by its two point function in $k$-space,
\begin{eqnarray}
\langle B_i^*(\vec{k})B_j(\vec{q})\rangle=(2\pi)^3\delta({\vec{k}-\vec{q}})P_B(k)\left(\delta_{ij}-\frac{k_ik_j}{k^2}\right),
\end{eqnarray}
where the power spectrum, $P_B(k)$ is given by \cite{kk11}
\begin{eqnarray}
P_B(k,k_m,k_L)=A_B\left(\frac{k}{k_L}\right)^{n_B}{\cal W}(k,k_m)
\end{eqnarray}
with $A_B$ its amplitude and $k_L$  a pivot wave number chosen to be 1 Mpc$^{-1}$ 
\;\,\footnote{
For {\sl helical}, gaussian random magnetic fields, in general, there is a second contribution to the 2-point function. For further details and the resulting CMB temperature anisotropies and polarization see, e.g., \cite{kk12}.}. 
Numerical solutions will be presented for negative spectral indices.
Moreover a gaussian window function ${\cal W}$ is used to implement effectively the damping of magnetic fields by photon viscosity in a process similar to the damping of density perturbations by photon diffusion or Silk damping before decoupling. This is encoded by 
an upper cut-off  comoving  wave number $k_m$ of the magnetic field spectrum by choosing
 ${\cal W}(k,k_m)=\pi^{-3/2}k_m^{-3}e^{-(k/k_m)^2}$.
 $k_m$ has its largest value at recombination  \cite{sb,jko} and is given for the bestfit parameters of Planck 2018 data only \cite{kk21,Planck-2018} by
\begin{eqnarray}
k_m=301.45\left(\frac{B}{\rm nG}\right)^{-1}{\rm Mpc}^{-1}.
\end{eqnarray}

\section{Multifrequency angular power spectra (MAPS)}
\label{s2}
\setcounter{equation}{0}

Observations of the 21 cm signal provide new possibilities to constrain cosmological parameters as well as, e.g., the thermal and ionisation history of the universe and in particular the epoch 
of reionization. In the rest frame of a neutral hydrogen atom the hyperfine transition of the ground state causes either emission or absorption of photons at a frequency $\nu_{21}=1420$ MHz corresponding to a wavelength $\lambda_{21}=21$ cm. The 21 cm line signal traces most importantly the distribution of neutral hydrogen in the universe at different redshifts along the line of sight (LOS).  
Interactions of CMB photons with HI along the LOS lead to a change in the CMB brightness temperature at a frequency $\nu=\nu_{21}/(z+1)$ and along the direction $\hat{n}$, $\delta T_{21}(\nu,\bf{\hat{n}})$. Depending on absorption or emission of photons by HI this is negative or positive.

Observing the integrated  21 cm line signal over a wide sky area underlies the HI intensity mapping (HI IM) technique. Within large enough 3D pixels it is expected to capture contributions from different HI regions yielding a strong signal. Using a specific spectral line, i.e. the 21 cm line, and observations at different frequencies allows to determine 
the corresponding redshifts of the HI regions with high precision. This yields an efficient way to survey the HI distribution in the universe allowing to probe large scale structure  (for more details and reviews, cf,  e.g. 
 \cite{2016ApJ.LiuZhangParsons,ska1-cosmo-redbook, LiuShaw2020PASP,carucci2020}).  Examples of 21 cm signal maps can be found, e.g., in \cite{LiuShaw2020PASP, PritchardLoeb}.
 Foreground cleaned MeerKAT intensity maps are shown in \cite{meerkat-IM}.

The 21 cm brightness temperature fluctuation $\delta T_{21}({\bf \hat{n}})$
 can be expanded in terms of spherical harmonics such that cf. e.g. \cite{datta2007,fg_rm,pal2208.11063}
 \begin{eqnarray}
\delta T_{21}(\hat{\bf n},\nu)=\sum_{\ell,m}a_{\ell m}(\nu)Y_{\ell}^m(\hat{\bf n})
\label{delT}
\end{eqnarray}
with the harmonic coefficients,
\begin{eqnarray}
a_{\ell m}=\int d{\bf\hat{n}}^2\delta T_{21}(\nu,{\bf\hat{n}})Y^*_{\ell m}({\bf\hat{n}}).
\label{a_ellm}
\end{eqnarray}

The frequency dependence means that the 21 cm brightness temperature fluctuation  $\delta T_{21}(\hat{\bf n},\nu)$   is no longer statistically homogeneous along the line of sight.
This property makes it  useful to define the multifrequency angular power spectrum (MAPS) as \cite{datta2007}
\begin{eqnarray}
C_{\ell}(\nu_1,\nu_2)=\langle a_{\ell m}(\nu_1)a_{\ell m}^*(\nu_2)\rangle.
\label{clnu1nu2}
\end{eqnarray}
However, considering small bandwidths     and assuming $|\Delta\nu|/\nu_1\ll 1$ with $\Delta\nu=\nu_2-\nu_1$ 
the signal along the LOS  is approximately statistically homogeneous.
Thus  the multifrequency angular power spectrum is a function of $\Delta\nu$ only  setting $\nu_1$ to a given reference frequency, say, $\nu_*$ such that \cite{datta2007, mondal2019,pal2208.11063},
\begin{eqnarray}
C_{\ell}(\Delta\nu,\nu_*)\equiv C_{\ell}(\nu_*,\nu_*+\Delta\nu).
\label{clDelnu}
\end{eqnarray} 
It is assumed that redshift dependent background quantities are calculated at the redshift corresponding to $\nu_*$, namely, $z_*=\nu_{21}/\nu_*-1$ with $\nu_{21}=1420$ MHz.
In particular,   results are calculated for
\begin{itemize} 
\item uGMRT Band 3 which has the central  frequency $\nu_c=432.8$ MHz corresponding to 21 cm line emission from HI regions at redshift $z=2.28$ \cite{elahiMNRAS529,pal2208.11063,elahi2301.06677}.
\item the MeerKAT L band configuration 
at central frequencies  (1)  $\nu_c=986$ MHz allowing HI intensity mapping at $z=0.44$ and (2) $\nu_c=1077.5$ MHz and HI regions at $z=0.32$ \cite{PaulMeerKAT2023}.
\end{itemize}
The future SKA1-MID covers a frequency range which includes all of these three central frequencies \cite{ska1-cosmo-redbook,SKAtelecon}. This is particularly interesting for the prospects of detection estimated by the signal-over-noise ratios which are presented in the following.
The 21 cm signal brightness temperature maps have been calculated at different frequencies  with  the  {\tt crime}  code \cite{crime} for a  bandwidth $\Delta\nu=10$ MHz over 106 frequency channels in the interval $[\nu_c,\nu_c+\Delta\nu]$ for each of the aforementioned central frequencies of uGMRT and MeerKAT. Cross correlating the 21 cm signal maps at different frequencies covers in total the range  $-10\leq\Delta\nu/({\rm MHz})\leq10$. The multifrequency angular power spectra (MAPS) $C_{\ell}(\nu_*,\Delta\nu)$ (cf. equations (\ref{clnu1nu2}) and (\ref{clDelnu}))
have been calculated with an adapted version of the \texttt{fg\_rm}  \cite{fg_rm} numerical code.  The 21 cm signal brightness temperature maps were simulated using  only the cosmological, theoretical model with the initial linear matter power spectrum determined by the adiabatic primordial curvature mode and the  compensated magnetic mode.

\begin{figure}[h!]
\centerline{\epsfxsize=3.7in\epsfbox{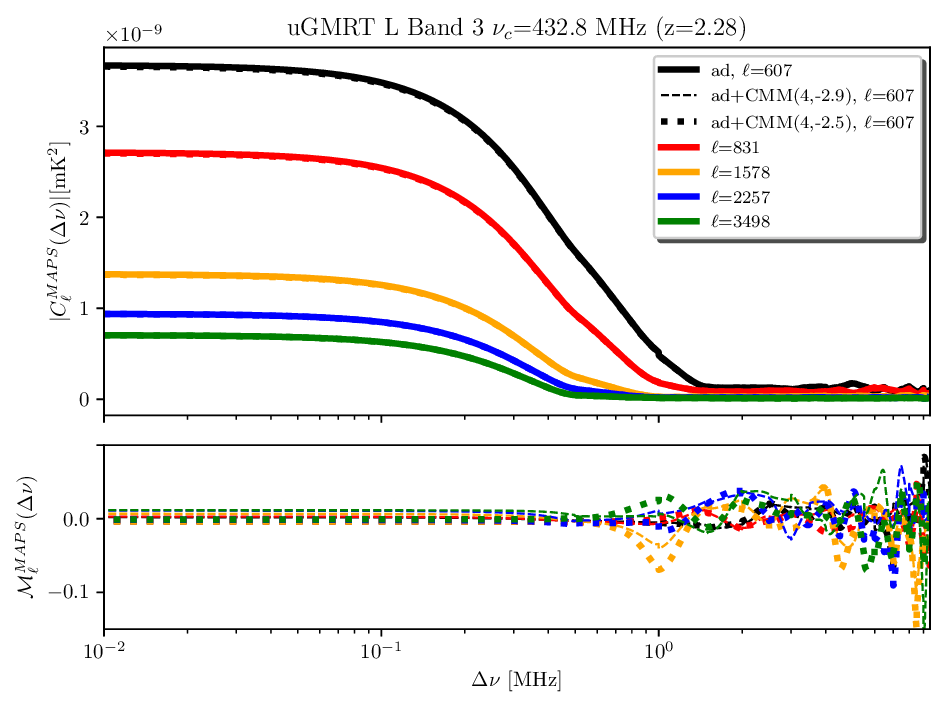}}
\centerline{\epsfxsize=3.7in\epsfbox{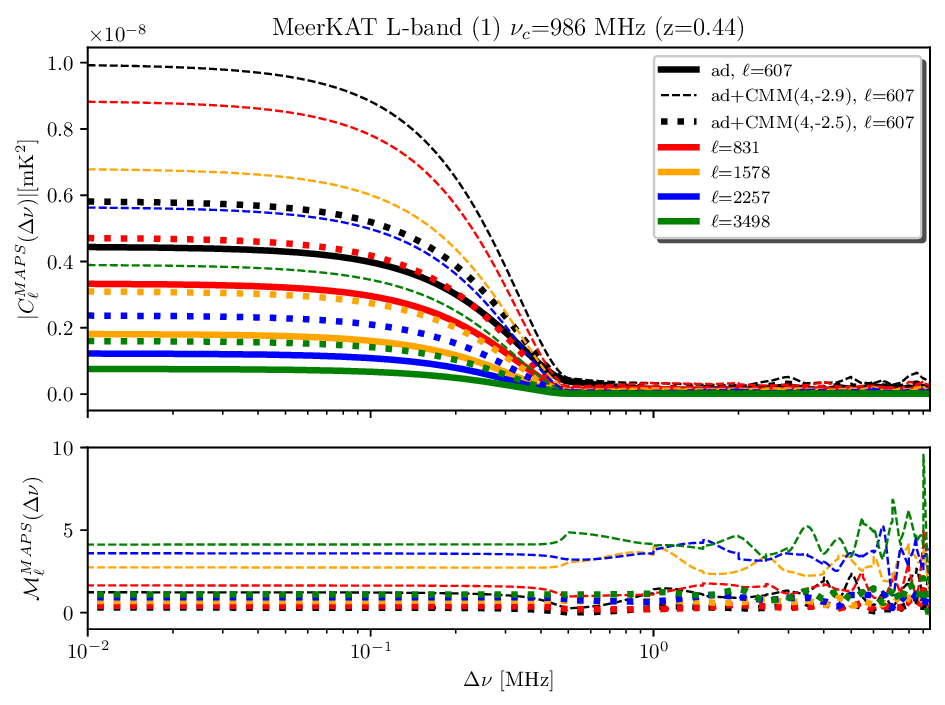}
\epsfxsize=3.7in\epsfbox{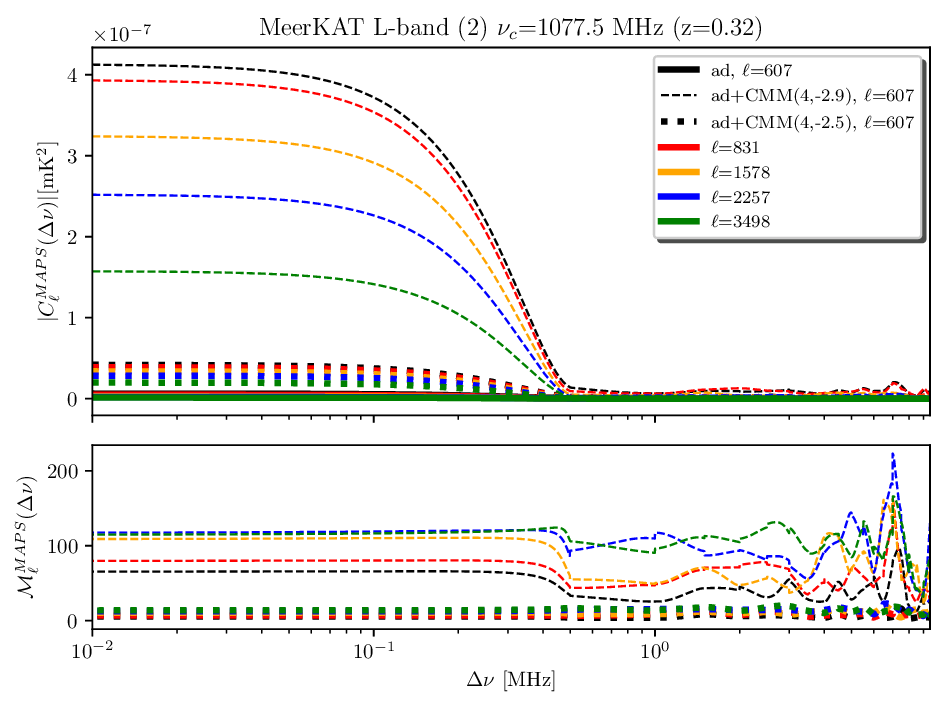}}
\caption{Absolute values of  the multi frequency angular power spectra of the theoretical model  21 cm signal  $C_{\ell}^{MAPS}(\Delta\nu)\equiv C_{\ell}(\nu_c,\Delta\nu)$ as a function of $\Delta\nu$ for the total  adiabatic  and compensated magnetic mode  are  shown for uGMRT  Band 3 at $\nu_c=432.8$ MHz ($z=2.28$)  ({\sl first row}), for MeerKAT L-band (1)  at  $\nu_c=986$ MHz ($z=0.44$) ({\sl second row, left})
and for  MeerKAT L-band (2)  at  $\nu_c=1077.5$ MHz ($z=0.32$) ({\sl second row, right}) in the upper panels. The fractional change of the  MAPS $ {\cal M}^{[ad,CMM(B_0,n_B)]}_{\ell}(\nu_{c,conf},\Delta\nu)$
resulting from the contribution of the adiabatic as well as the magnetic mode with respect to the MAPS of the adiabatic mode only are reported in the corresponding lower panels.
These are shown  for different magnetic field parameters ($B_0=0$ ({\sl solid}), $B_0=4$nG: $n_B=-2.9$ ({\sl light dashed}), $n_B=-2.5$ ({\sl thick dashed})).
Cosmological parameters are chosen to be the best fit parameter of the Planck 2018 
baseline $\Lambda$CDM model.
}
\label{fig0a}
\end{figure}

Since $|\Delta\nu|\ll \nu_{c, conf}$, where $conf$ stands for the chosen radio telescope configuration, uGMRT Band 3, or MeerKAT L  band configuration (1) or (2), respectively, the evolution of the background cosmological variables can be ignored approximately. 
Therefore the MAPS  to a very good approximation depend only on the modulus $|\Delta\nu|$ leading to $C_{\ell}(\nu_{c,conf},|\Delta\nu|)$.  
Thus in the final numerical results only $\Delta\nu\geq 0$ has to be taken into account.
Moreover, $\Delta\nu=0$ corresponds to the auto frequency correlation function of the 21 cm signal brightness temperature fluctuations.
Finally, for each radio telescope array configuration the MAPS   are calculated by averaging over all values at a given   $|\Delta\nu|$.  
Using an interpolation routine of the {\tt Python SciPy} mathematical algorithms \cite{2020SciPy-NMeth} the MAPS can be 
numerically calculated as a function of $\Delta\nu$ for fixed multipole $\ell$ as shown in figure \ref{fig0a} for different choices of the magnetic field parameters 
and radio telescope array configurations. 
For results of MAPS $C_{\ell}(\Delta\nu)$ for the uGMRT Band 3 data see, e.g., \cite{pal2208.11063}. Here results have been shown at fixed multipoles with values as used 
in the analysis of the uGMRT Band 3 data \cite{elahi2301.06677}.
Moreover, as way of example numerical solutions are presented for a magnetic field with amplitude $B_0=4$ nG and spectral indices $n_B=-2.9$ and $n_B=-2.5$, respectively.

The effect of choosing different magnetic field parameters can be captured by considering the difference between the resulting MAPS.
Therefore, the 
fractional change between the modulus of the  MAPS of the combined effect of the adiabatic primordial curvature mode, denoted by $ad$,  and the compensated magnetic mode $CMM(B_0,n_B)$
w.r.t. to the one taking into account the adiabatic mode only is defined by
\begin{eqnarray}
{\cal M}^{[ad,CMM(B_0,n_B)]}_{\ell}(\nu_{c,conf},\Delta\nu)=\frac{|C_{\ell}^{[ad+CMM(B_0,n_B)]}(\nu_{c,conf},\Delta\nu)|-|C_{\ell}^{[ad]}(\nu_{c,conf},\Delta\nu)|}{|C^{[ad]}_{\ell}(\nu_{c,conf},\Delta\nu)|}
\end{eqnarray}
Fractional changes w.r.t. to the adiabatic mode are shown   as a function of $\Delta\nu$ for fixed multipole $\ell$ in figure \ref{fig0a} for different choices of the magnetic field parameters and radio telescope array configurations. 

A significant decrease in amplitude for $\Delta\nu>1$ MHz
can be observed in figure \ref{fig0a} from the multifrequency angular power spectra $C_{\ell}(\nu_{c,conf},|\Delta\nu|)$.
This manifests that  the frequency cross correlated 21 cm signal decorrelates rapidly with growing difference between frequencies, $\Delta\nu$. 
The rate of decorrelation depends on $\ell$ as well. It is slower on larger scales, that is for smaller $\ell$. 
In general the decorrelation of the 21 cm signal across different frequencies depends on the evolution of the physical processes in the HI regions along the LOS determining the 21 cm signal
\cite{shawAK2023mnras522}.

\subsection{Signal-over-noise ratio}
The signal-over-noise ratio provides a good estimate of the prospects of detection. It is assumed that foregrounds have been cleaned and thus only system noise will be included.
This  is estimated using the noise power spectrum given by \cite{Alvarez:2005sa,Adshead:2007ij,Zaldarriaga:2003du,kk23},
\begin{eqnarray}
\frac{\ell^2  {\cal N}_{\ell}^{21}}{2\pi}=\frac{(130 \mu K)^2}{N_{month}\Delta\nu^{BW}_{\rm MHz}}
\left[
\left(\frac{\ell}{100}\right)
\left(\frac{1+z}{10}\right)\left(\frac{D}{1{\rm km}}\right)\left(\frac{10^3 {\rm m}^2{\rm K}^{-1}}
{A_{eff}/T_{sys}}\right)
\right]^2,
\label{Nell}
\end{eqnarray}
where $N_{month}$[months] is the total observation time. 
Denoting in brackets $(\Delta\nu^{BW}_{\rm MHz}, D, A_{eff}/T_{sys})$ with $\Delta\nu^{BW}_{\rm MHz}[{\rm MHz}]$  the bandwidth,  $D[{\rm km}]$  the baseline  and $A_{eff}/T_{sys}\; [{\rm m^2/K}]$  the ratio of the effective area over system temperature the following technical specifications are used for the different radio telescope arrays \cite{SKAtelecon}: (u)GMRT(450,27,250),  MeerKAT(1000,4,321) and  SKA1-MID(770,150,1560).

The maximal multipole $\ell_{max}$ which could be detected at a given baseline length is given by \cite{Alvarez:2005sa, Zaldarriaga:2003du}
\begin{eqnarray}
\ell_{max}=2\pi\frac{D}{\lambda}=2994\left(\frac{D}{1{\rm km}}\right)\left(\frac{10}{1+z}\right)
\end{eqnarray}
for $\lambda=(1+z) 21 {\rm cm}$. This yields $\ell_{max}$ to be $2.5\times 10^5$ (uGMRT), $8.3\times 10^4$ (MeerKAT L-band (1)), $9.1\times 10^4$ (MeerKAT L-band (2)).
For  SKA1-MID  $\ell_{max}$ is given by $1.4\times 10^6$,  $3.1\times 10^6$ and $3.4\times 10^6$, respectively, at $z=2.28$, $z=0.44$ and $z=0.32$, respectively.

The signal-over-noise ratio of the frequency cross correlation angular power spectrum $C_{\ell}(\nu_c,\Delta\nu)$ can be estimated using, e.g., \cite{Ma,Adshead:2007ij,Dore,kk23} (see also 
\cite{shawAK2023mnras522,mondal2020mnras494})
by
\begin{eqnarray}
\left(
\frac{S}{N}\right)_{\ell}^2=\frac{f_{\rm sky}(2\ell+1)\Delta\ell_{\rm bin}C_{\ell}(\Delta\nu)^2}{\left( C_{\ell}(\nu_c,\nu_c)+{\cal N}_{\ell}^{21}\right)^2+\left( C_{\ell}(\Delta\nu)\right)^2},
\end{eqnarray}
where $\Delta\ell_{\rm bin}\simeq 0.46\ell$ is the bin width at a given $\ell$ and $f_{\rm sky}$ the observed sky fraction. 
The latter is given in terms of the solid angle covering the field of observations \cite{Alvarez:2005sa}
\begin{eqnarray}
f_{\rm sky}\equiv\frac{\Omega}{4\pi}=2.424\times 10^{-3}\left(\frac{\Omega}{100 {\rm\, deg}^2}\right).
\end{eqnarray}

\begin{figure}
\centerline{\epsfxsize=2.8in\epsfbox{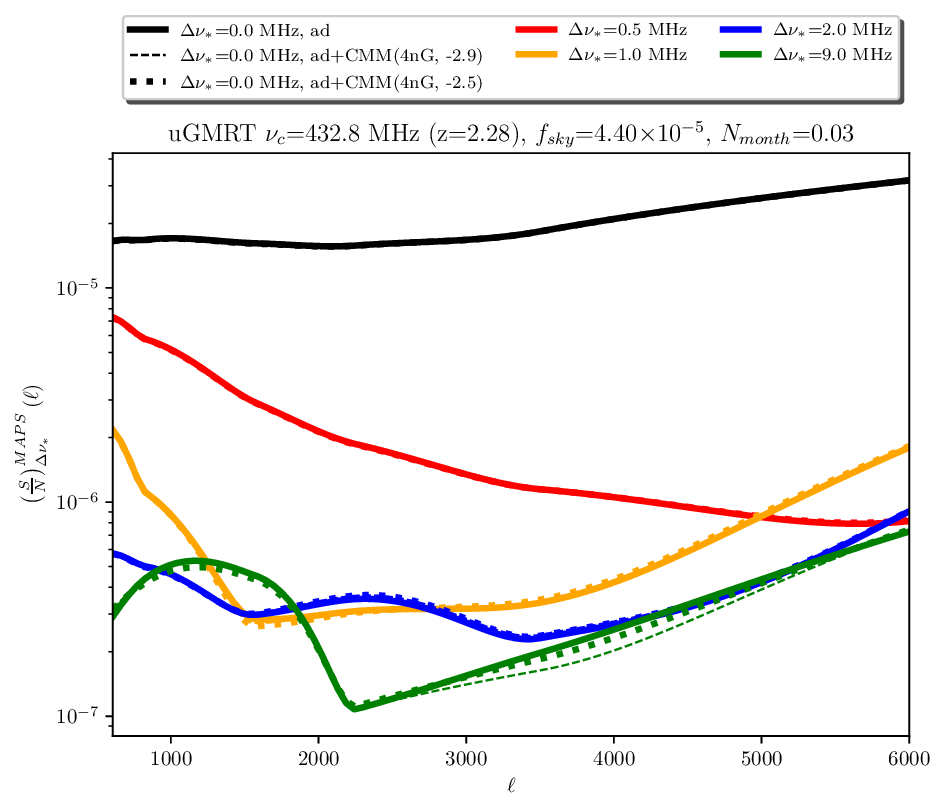}
\epsfxsize=2.8in\epsfbox{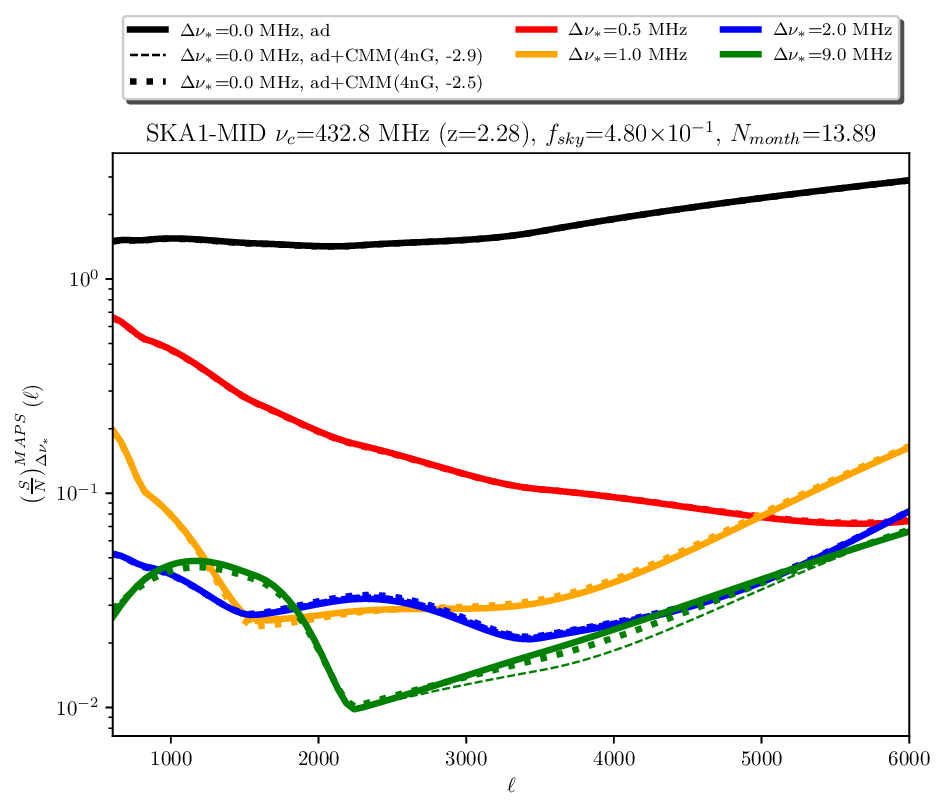}}
\centerline{\epsfxsize=2.8in\epsfbox{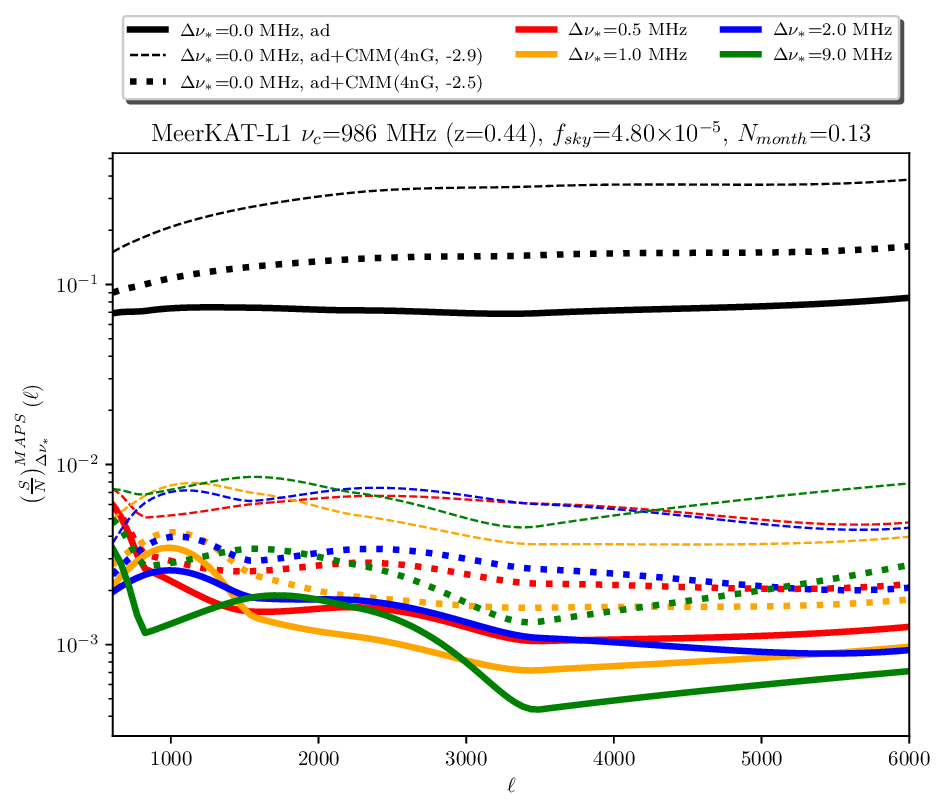}
\epsfxsize=2.8in\epsfbox{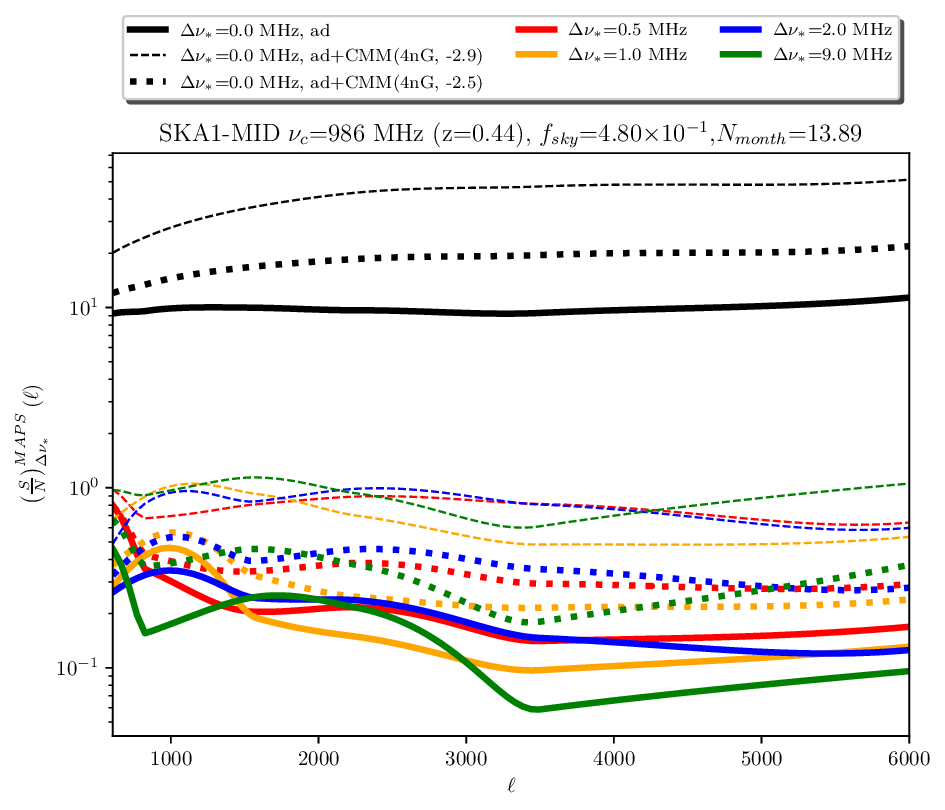}}
\centerline{\epsfxsize=2.8in\epsfbox{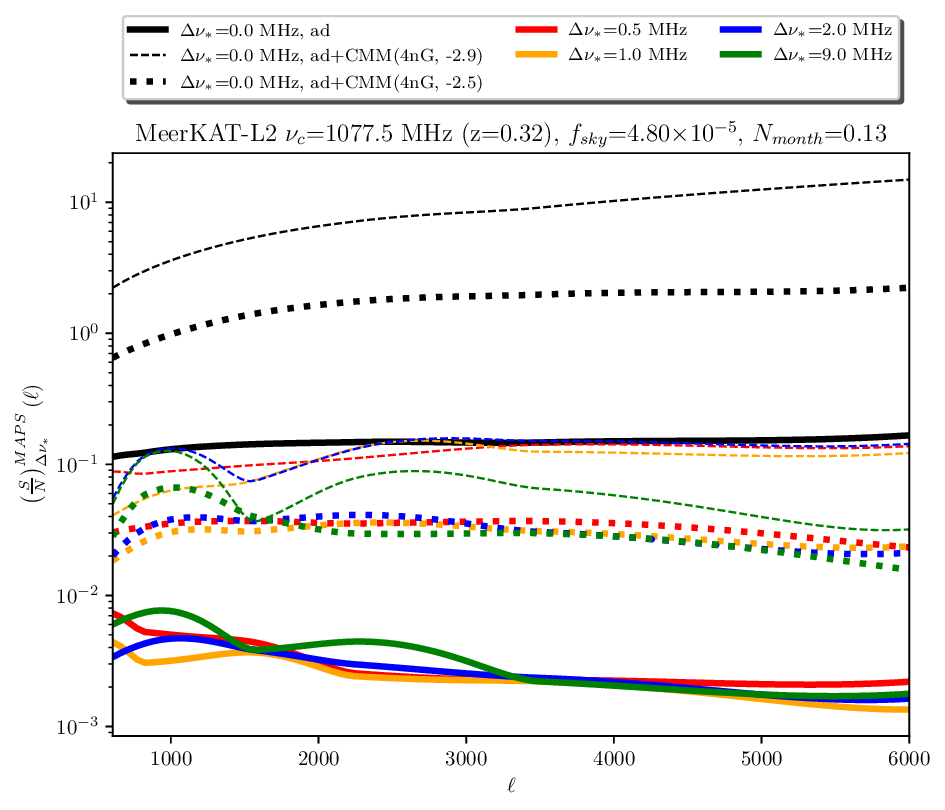}
\epsfxsize=2.8in\epsfbox{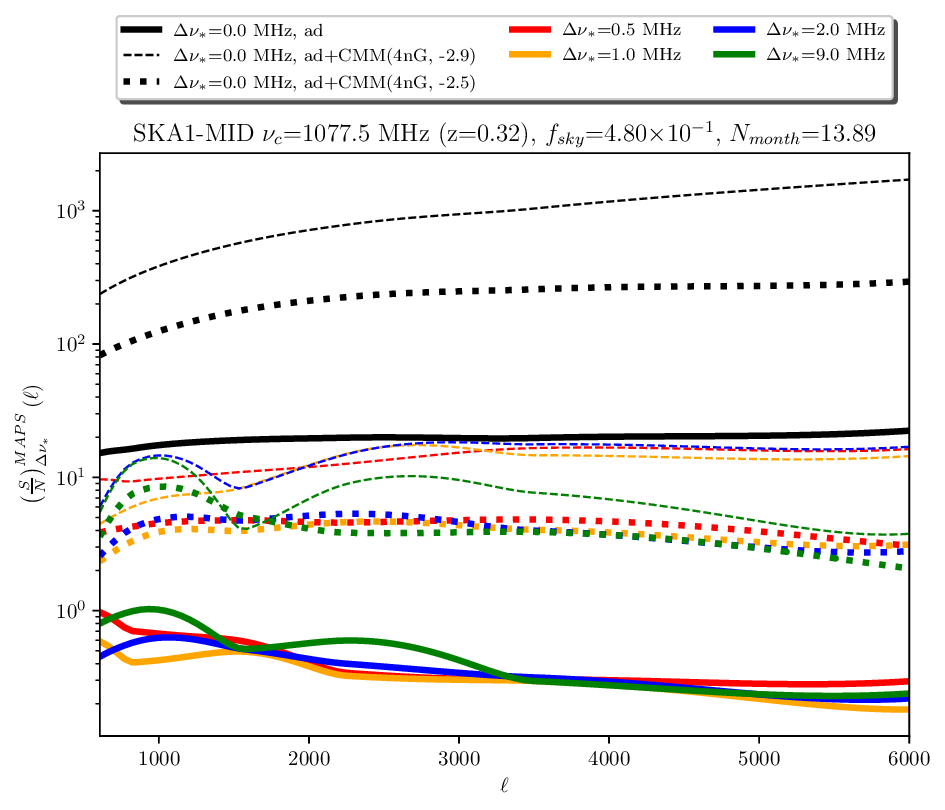}}
\caption{Signal-over-noise ratios of  the multi frequency angular power spectra $C_{\ell}^{MAPS}(\Delta\nu)\equiv C_{\ell}(\nu_c,\Delta\nu)$  at different values of $\Delta\nu=\Delta\nu_*$ for the total  adiabatic  and compensated magnetic mode  for ({\sl left column, top to bottom}),  uGMRT  Band 3 at $\nu_c=432.8$ MHz ($z=2.28$) , for MeerKAT L-band (1)  at  $\nu_c=986$ MHz ($z=0.44$)  and for  MeerKAT L-band (2)  at  $\nu_c=1077.5$ MHz ($z=0.32$). 
({\sl Right column, top to bottom}) shows results for the proposed SKA1-MID Wide Band 1 Survey at central frequencies for uGMRT and MeerKAT L-band (1) and (2), respectively.
These are shown  for different magnetic field parameters ($B_0$=0 ({\sl solid}), $B_0=4$nG: $n_B=-2.9$ ({\sl light dashed}), $n_B=-2.5$ ({\sl thick dashed})).
Further technical characteristics are given in the text.}
\label{fig2}
\end{figure}

Moreover it was used that noise across different frequency channels is not correlated. Furthermore since $\Delta\nu/\nu_c\ll 1$ the auto correlation angular power spectrum is approximated by $C_{\ell}(\nu_c+\Delta\nu)\simeq C_{\ell}(\nu_c)$ and similarly for the noise power spectrum.
In figure \ref{fig2} the signal-over-noise ratios are shown for current characteristics of observations of uGMRT, namely $\Omega$=1.8 deg$^2$  and observation time $t_{\rm obs}=$25 h \cite{elahiMNRAS529}, and $\Omega$=2.0 deg$^2$ and  $t_{\rm obs}=$96 h for MeerKAT L-Band (1) and (2), respectively, \cite{PaulMeerKAT2023}.
Moreover, S/N is  reported for the proposed future Wide Band 1 Survey with SKA1-MID at the corresponding central frequencies of uGMRT and MeerKAT L-Band (1) and (2). The Wide Band 1 Survey  is proposed  to cover an area of $\Omega$=20000 deg$^2$ on the sky and $t_{\rm obs}=$10000 h \cite{ska1-cosmo-redbook}.
Assuming perfect foreground cleaning
 $S/N$ of ${\cal O}(1)$ could be reached at $\Delta\nu=0$ for a contributing compensated magnetic mode with spectral index $n_B=-2.9$ and growing with multipole $\ell$ for the current observational characteristics of the MeerKAT L band (2) configuration (cf.  figure \ref{fig2} ({\sl left column})). In figure \ref{fig2}  ({\sl right column}) results are shown for the future SKA1-MID for the proposed Wide Band 1 
 survey at the corresponding frequencies of the central frequencies used for uGMRT and MeerKAT L band (1) and (2), respectively. Signal-over-noise ratios are enhanced due to the improved technical characteristics of SKAO as well as a much longer total observation time and  larger sky area that will be covered by the Wide Band 1 Survey.
In this case at all three central frequencies, namely,  432.8 MHz, 986 MHz and 1077.5 MHz, at least the auto correlation power spectra of the 21 cm line signal, i.e. $\Delta\nu=0$, would in principle be observable. Moreover, at the MeerKAT L band (2) central frequency $\nu_c=1077.5$ MHz multifrequency angular power spectra for $\Delta\nu>0$ reach $S/N> 1$  with the possibility to  constrain the  magnetic field parameters.
Signal-over-noise ratios for 21 cm multifrequency angular power spectra MAPS have been obtained for SKA1-low , e.g., \cite{mondal2020}.

\begin{figure}
\centerline{\epsfxsize=2.8in\epsfbox{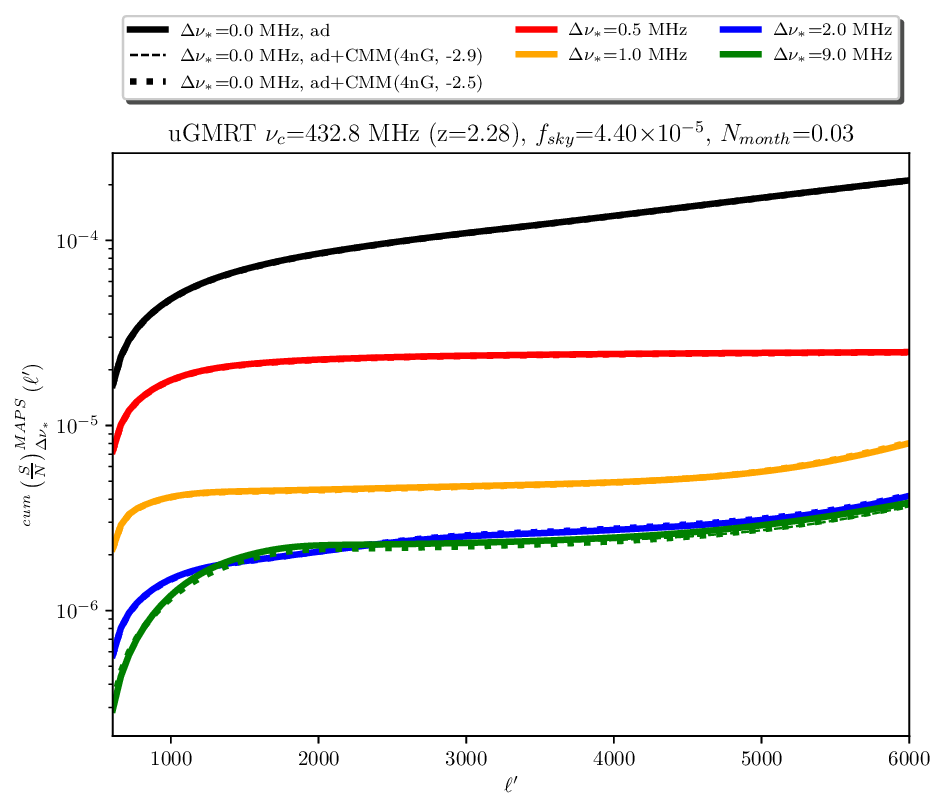}
\epsfxsize=2.8in\epsfbox{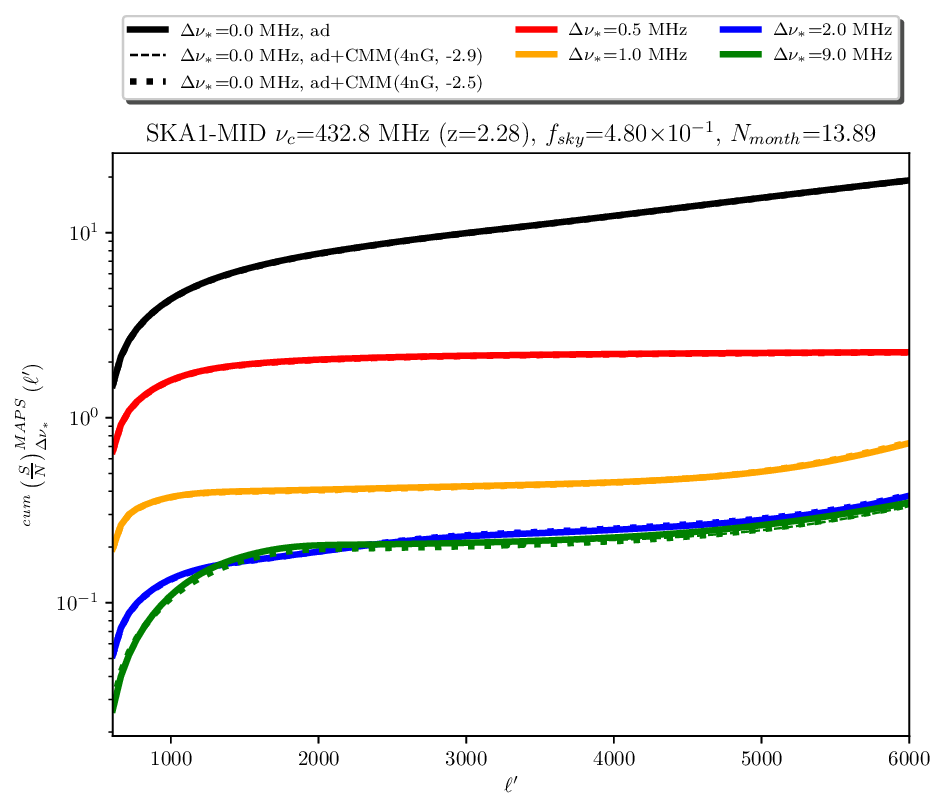}}
\centerline{\epsfxsize=2.8in\epsfbox{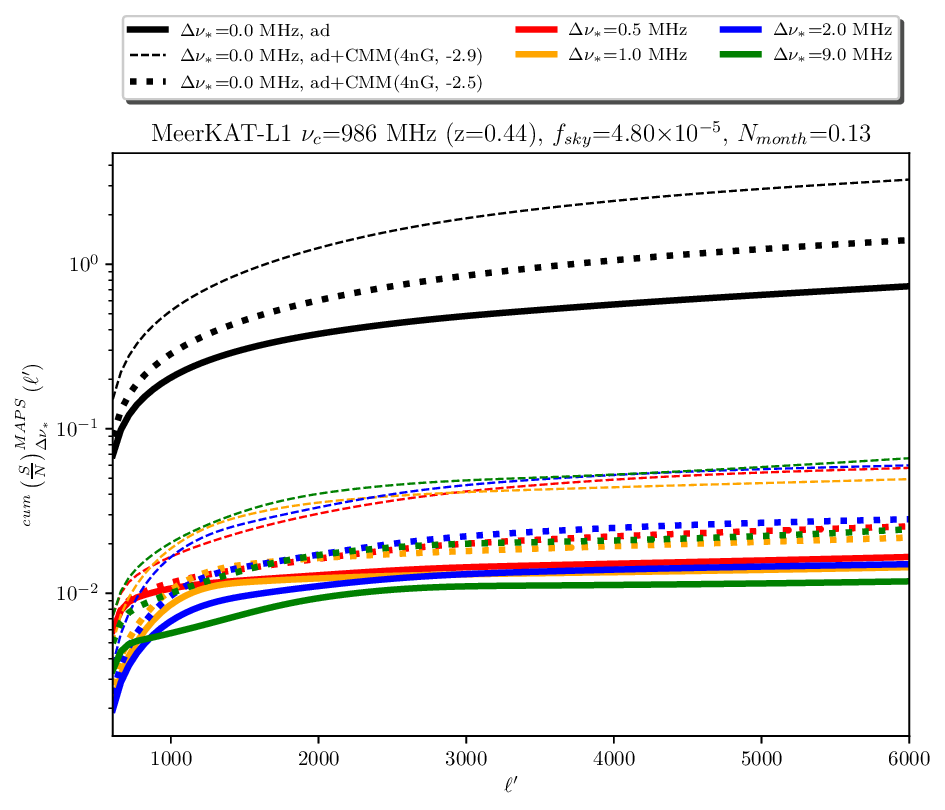}
\epsfxsize=2.8in\epsfbox{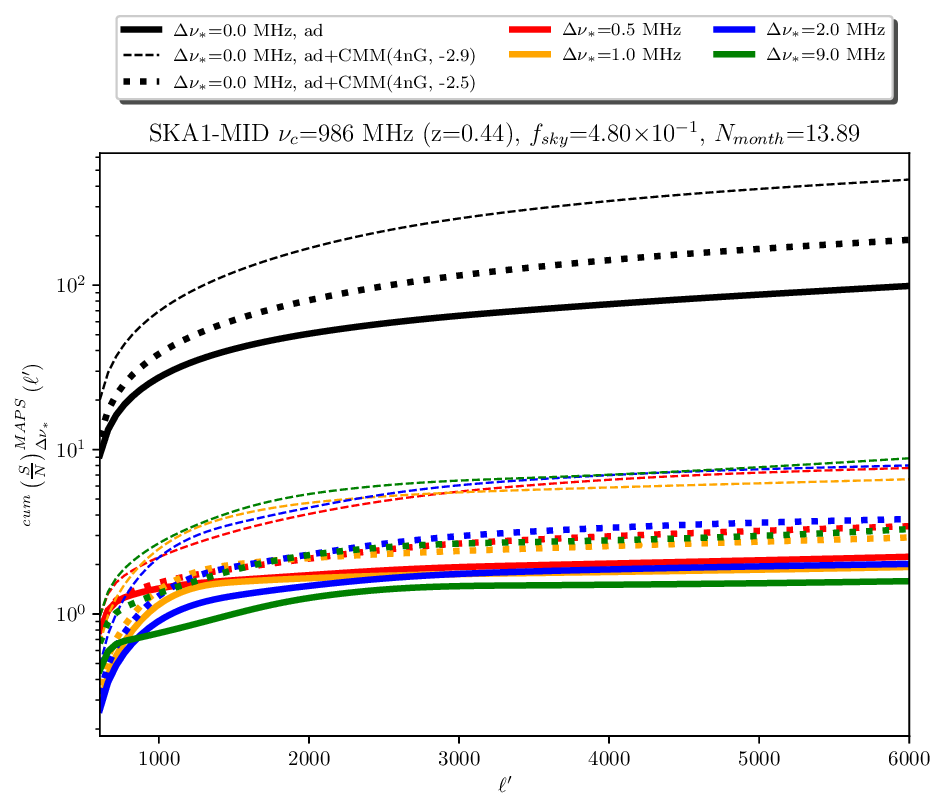}}
\centerline{\epsfxsize=2.8in\epsfbox{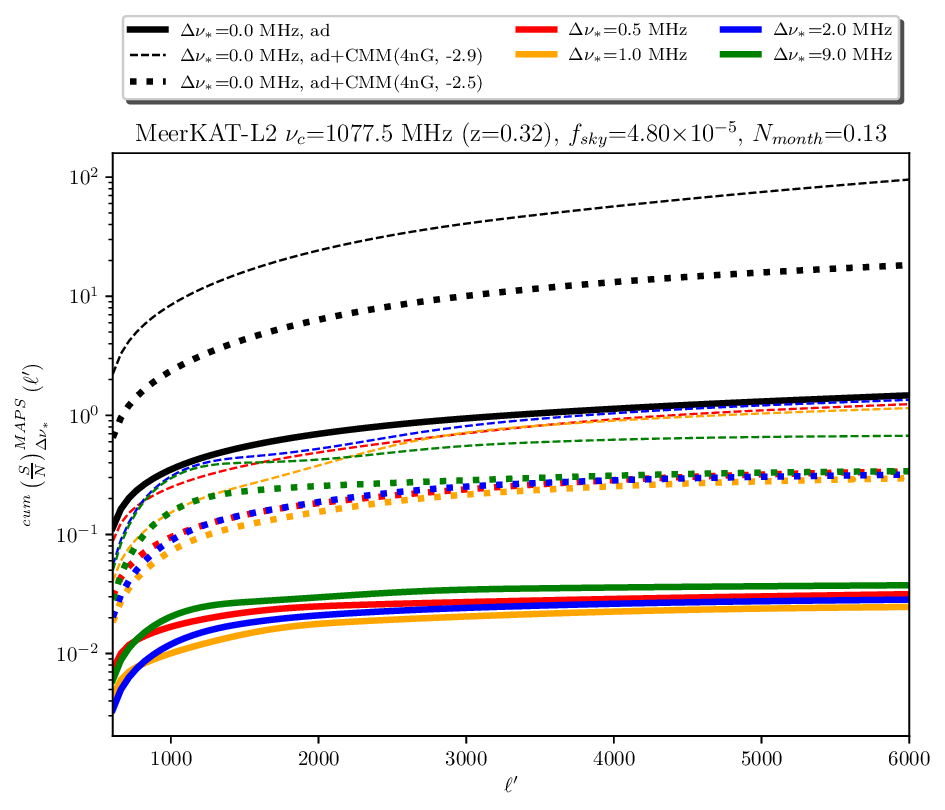}
\epsfxsize=2.8in\epsfbox{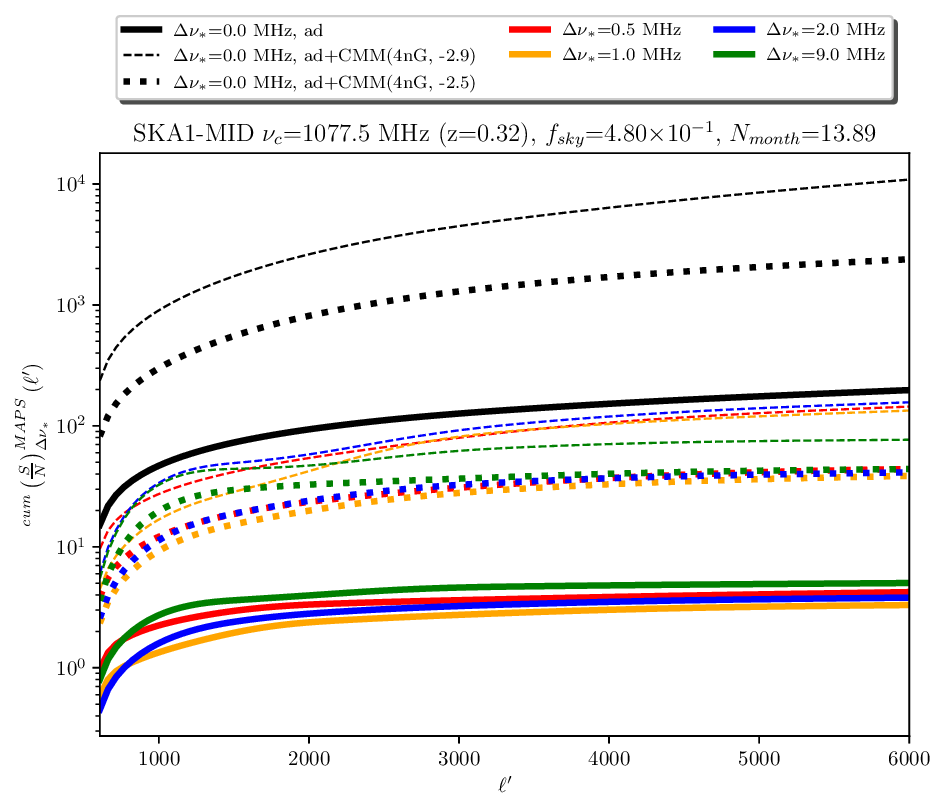}}
\caption{Cumulative signal-over-noise ratios of  the multi frequency angular power spectra $C_{\ell}^{MAPS}(\Delta\nu)\equiv C_{\ell}(\nu_c,\Delta\nu)$  at different values of $\Delta\nu=\Delta\nu_*$ for the total  adiabatic  and compensated magnetic mode  for ({\sl left column, top to bottom}),  uGMRT  Band 3 at $\nu_c=432.8$ MHz ($z=2.28$) , for MeerKAT L-band (1)  at  $\nu_c=986$ MHz ($z=0.44$) and for  MeerKAT L-band (2)  at  $\nu_c=1077.5$ MHz ($z=0.32$). 
({\sl Right column, top to bottom}) shows results for the proposed SKA1-MID Wide Band 1 Survey at central frequencies for uGMRT and MeerKAT L-band (1) and (2), respectively.
These are shown  for different magnetic field parameters ($B_0$=0 ({\sl solid}), $B_0=4$nG: $n_B=-2.9$ ({\sl light dashed}), $n_B=-2.5$ ({\sl thick dashed})).
Further technical characteristics are given in the text.}
\label{fig3}
\end{figure}

In general the cumulative signal-over-noise ratio upto a given multipole $\ell'$ 
is determined by \cite{Ma}
\begin{eqnarray}
^{\rm cum}\left(\frac{S}{N}\right)^2(< \ell')=\sum_i \left(
\frac{S}{N}\right)_{\ell_i}^2
\end{eqnarray}
where $i$ denotes the $ith$ bin with central multipole $\ell_i$ and $\ell '$ the maximal multipole implying $\ell_i\leq \ell '$.
By definition $(S/N)_{cum}(\ell')$ yields the total $(S/N)$ upto $\ell'$.
In \cite{Ma}  the cumulative $S/N$ was calculated for  cross correlations of the  kinetic Sunyaev-Zel'dovich effect (kSZ)  squared field and 21cm signals.
For cross correlations between the CMB Doppler mode and 21 cm line signal in the presence of primordial magnetic fields examples of $(S/N)_{cum}$ can be found in \cite{kk23}.
The cumulative signal-over-noise ratios for uGMRT, MeerKAT and the SKA1-MID are shown in figure \ref{fig3}.
As can be appreciated in figure \ref{fig3} ({\sl left column})  the resulting 
 $^{\rm cum}\left(\frac{S}{N}\right)(< \ell')$ become larger than 1 for $\Delta\nu=0$ for growing multipole for magnetic spectral indices $n_B=-2.9$ for the MeerKAT L band (1) and (2), respectively, current observational specifications as well as for the latter for $n_B=-2.5$.
In figure \ref{fig3} ({\sl right column}) the corresponding results for  the future SKA1-MID for the proposed Wide Band 1 
 survey are presented. The cumulative signal-over-ratios reach values larger than 1 for MAPS ($\Delta\nu\geq 0$) at all three reference frequencies: at the uGMRT reference frequency upto $\Delta\nu=0.5$ MHz. For the MeerKAT L band (1) reference frequency for larger multipoles it reaches values at least ${\cal O}(1)$ and for MeerKAT L band (2) amplitudes are increasing for all frequency separations $\Delta\nu\leq 9$ MHz under consideration. 
 At larger multipoles it levels off indicating that going to much larger multipoles might not improve significantly the signal-over-noise ratio.
 In general prospects for constraining magnetic field parameters are improving for smaller frequency separations $\Delta\nu$ and larger multipoles.
However, this is only a first study of the effects of a primordial magnetic field on the 21 cm multifrequency angular power spectrum and 
its possible detection. It only uses the cosmological, theoretical model signal and approximates the noise power angular power spectrum (cf equation (\ref{Nell})).
More realistic estimates of $S/N$ would be obtained, e.g., by improved foreground modeling  (cf e.g. \cite{elahi2301.06677}), more precise estimates of telescope sensitivities (using e.g. the Python package {\tt 21cmSense} \cite{21cmSense})  and possible mode filtering (cf. e.g. \cite{trottMWA}). Moreover to constrain the magnetic field parameters a full numerical parameter estimation needs to be done (using e.g. {\tt 21CMMC}) \cite{21CMMC}). However, this is beyond the scope of this article and will be part of future work.

\section{Conclusions}
\label{sec3}
\setcounter{equation}{0}
Observations of the cosmic 21 cm brightness temperature fluctuations allow to study physical conditions along the evolution of the universe, and in particular the epoch of reionization. 
This allows to constrain cosmological parameters. Here prospects of constraining parameters of a putative primordial, cosmological magnetic field have been considered.
Primordial magnetic fields present before recombination influence the matter power spectrum which has been used in the simulation of 21 cm signal temperature maps for different magnetic field parameters. As the signal  originates in regions of neutral hydrogen at different redshifts it is observed at different frequencies. Cross correlating temperature maps at different frequencies allows to calculate multifrequency angular power spectra (MAPS). These have been obtained for specifications for current observations with  the  uGMRT Band 3 as well as the MeerKAT L band. Assuming perfect foreground removal signal-over-noise ratios and cumulative signal-over-noise ratios have been calculated including the system noise of  these radio  telescope arrays as well as for the specifications of the future SKA1-MID and  the SKA1-MID proposed Wide Band 1 Survey. In particular for the latter signal-over-noise ratios larger than one can be obtained even for larger frequency separations for larger multipoles. Moreover, the results  in this first study seem to be promising to constrain parameters of a primordial magnetic field with multifrequency angular power spectra using 21 cm line intensity mapping observations.

However, in future work, there are several aspects that should be taken into account to obtain more complete, realistic estimates of primordial magnetic field parameters.
These include the effect of additional heating by magnetic field decay  using MHD simulations.
It might also be interesting to use multifrequency cross correlations
to probe the evolution of primordial magnetic fields, cf. \cite{kk22}.
A full numerical cosmological parameter estimation using data will be done in the future.
There are also additional sources of uncertainties that should be taken into account by  improved foreground modeling  (cf e.g. \cite{elahi2301.06677}), more precise estimates of telescope sensitivities (using e.g. the Python package {\tt 21cmSense} \cite{21cmSense})  or mode filtering (cf. e.g. \cite{trottMWA})


\section{Acknowledgements}

Financial support by Spanish Science Ministry grants PID2024-160856NB-I00,
PID2021-123703NB-C22 (MCIU/AEI/FEDER, EU) and Basque Government grant IT1628-22 is gratefully acknowledged.


\appendix
\section{Foregrounds emissions and removal}
\label{app}
Foreground emissions are both of galactic, namely, synchrotron and free-free emission, and extragalactic origins  
such as from bright radio galaxies \cite{fg_rm}. 
Removal of these foreground contributions to the observed signal is one of the key objectives to isolate the cosmic 21 cm line signal. 

There are different foreground removal techniques, so-called blind or non-blind, depending on whether only generic foreground characteristics are used or specific models.
To clean the 21 cm intensity maps of the foregrounds pipelines have been and are being developed for current and future instruments.

In \cite{fg_rm} the  numerical code \texttt{fg\_rm}  
was presented to remove the foregrounds from the observed 21 cm intensity maps.
It uses blind foreground removal applying different methods such as principal component analysis, independent component analysis and line-of-sight polynomial fitting.


\bibliography{references}

\bibliographystyle{apsrev}

\end{document}